\documentclass[aps,pra,twocolumn,superscriptaddress,longbibliography]{revtex4-1}

\usepackage{natbib}
\usepackage{pgfplots}
\usepackage[breaklinks]{hyperref}
\usepackage{amssymb}
\usepackage{amsmath,amsthm}
\usepackage{amsfonts,dsfont}
\usepackage{graphicx}
\usepackage{lipsum}
\usepackage{subcaption}
\usepackage{mathtools}
\usepackage{tikz}
\usepackage{cleveref}
\usetikzlibrary{positioning}
\usepackage{color, colortbl}
\definecolor{Gray}{gray}{0.9}

\usepackage[normalem]{ulem}
\usepackage{booktabs}
\usepackage{xcolor}

\newcommand{\rr}{{\mathbf r}}

\newcommand{\pp}{\mathbf{p}}

\newcommand{\vt}{\mathbf{t}}
\newcommand{\vs}{\mathbf{s}}

\newcommand{\qq}{\mathbf{q}}
\newcommand{\kk}{\mathbf{k}}

\newcommand{\qn}{\tilde{q}}
\newcommand{\wn}{\tilde{\omega}}
\newcommand{\kn}{\tilde{k}}
\newcommand{\qqn}{\tilde{\mathbf{q}}}
\newcommand{\kkn}{\tilde{\mathbf{k}}}

\begin{document}

\title{Homogeneous Electron Liquid in Arbitrary Dimensions: \\ Exchange and Correlation Using the Singwi-Tosi-Land-Sj\"olander Approach}

\author{L. V. Duc Pham}
\affiliation{Institut f\"ur Physik, Martin-Luther-Universit\"at
Halle-Wittenberg, 06120 Halle (Saale), Germany}

\author{Pascal Sattler}
\affiliation{Institut f\"ur Physik, Martin-Luther-Universit\"at
Halle-Wittenberg, 06120 Halle (Saale), Germany}

\author{Miguel A. L. Marques}
\affiliation{Institut f\"ur Physik, Martin-Luther-Universit\"at
Halle-Wittenberg, 06120 Halle (Saale), Germany}

\author{Carlos L. Benavides-Riveros} 
\email{cl.benavidesriveros@unitn.it}
\affiliation{Pitaevskii BEC Center, CNR-INO and Dipartimento di Fisica, Università di Trento, I-38123 Trento, Italy}
\affiliation{Max Planck Institute for the Physics of Complex Systems,  N\"othnitzer Str.~38, 01187, Dresden, Germany}

\date{\today}

\begin{abstract}
 The ground states of the homogeneous electron gas and the homogeneous electron liquid are cornerstones in quantum physics and che\-mis\-try.  They are archetypal systems in the regime of slowly varying densities in which the exchange-corre\-la\-tion energy can be estimated with a myriad of methods. For high densities, the behaviour the energy is well-known for 1, 2, and 3 dimensions. Here, we extend this model to arbitrary integer dimensions, and compute its correlation energy beyond the random phase approximation (RPA), using the celebrated approach developed by Singwi, Tosi, Land, and Sjölander (STLS), which is known to be remarkably accurate in the description of the full electronic density response for $2D$ and $3D$, both in the paramagnetic and ferromagnetic ground states. For higher dimensions, we compare the results obtained for the correlation energy using the STLS method with the values previously obtained using RPA. We found that at high dimensions STLS tends to be more physical in the sense that the infamous sum rules are better satisfied by the theory. We furthermore illustrate the importance of the plasmon contribution to STLS theory.
\end{abstract}

\maketitle

\section{Introduction}

 The ground states of the homo\-ge\-neous electron gas (HEG) and the homogeneous electron liquid (HEL) have played a prominent role in the modelling  and understanding of a wide range of interacting electronic systems~\cite{Sommerfeld1928,PhysRevB.15.2819,PhysRevB.66.235116, PhysRevLett.105.086403,PhysRevLett.107.110402,PhysRevB.85.081103,JEP_2018__5__79_0,PeterReview,PhysRevLett.100.236404,PhysRevA.99.052501,PhysRevLett.102.126402,PhysRevB.88.085121}. These systems are some of the most important models of choice to develop, improve and benchmark many approximate approaches to the full ma\-ny-electron problem, including some of the most popular exchange-correlation functionals of density functional theory (DFT)~\cite{PhysRev.165.18,PhysRevLett.45.566,Jones15,PhysRevLett.100.036401}. 
 
An important question about these models is the dependency of the correlation energy, and other physical quantities, on $D$, the  dimension of the physical space in  which the gas/liquid is embedded. Indeed, there is a wealth of experimental elec\-tro\-nic setups in which one or two of the physical dimensions are much smaller than the remaining ones. They can thus be modeled as one- or two-di\-men\-sional quantum systems \cite{Wagner2012,PhysRevA.93.021605,Schmidt2019,Constantin,Gedeon_2022}. Furthermore, reduced dimensional systems often exhibit notable physical properties, ranging from Luttinger physics~\cite{Voit_1995} to Moir\'e superlattices~\cite{Kennes2021}. More recently, progress in the fabrication of artificial materials is paving the way for the realization of non-integer dimensions, as fractal substrates  (e.g., Sierpi\'nski carpets of bulk Cu) confining electron gases  \cite{Kempes,PhysRevB.93.115428,PhysRevResearch.2.013044}. The possibility to circumvent the von Neumann-Wigner theorem or to produce unconventional topological phases by engineering (or mimicking) additional synthetic dimensions \cite{PhysRevLett.127.023003,Kanungo2022,PhysRevA.95.063612,PhysRevLett.112.043001} also highlights the importance of realizing, studying and understanding interacting fermionic and electronic systems embedded in non-conventional dimensions.

For the high-density spin-unpolarized HEG in $D = 1, 2$, and 3, the energy per electron can be expanded in terms of $r_s$, the Wigner-Seitz radius, as follows \cite{giuliani_vignale_2005}:
\begin{align}
\label{formula1}
\varepsilon_D(r_s\rightarrow0) = \frac{a_D}{r_s^2} - \frac{b_D}{r_s} + c_D \ln r_s + \mathcal{O}(r_s^0) \,.
\end{align}
 The constants $a_D$, $b_D$ and $c_D$ are independent of $r_s$, and their functional form in terms of the dimension $D$ is well known \cite{PeterReview}. Quite remarkably, the logarithm that appears in Eq.~\eqref{formula1} for the 3D case is due to the long range of the Coulomb repulsion, and cannot be obtained from standard second-order perturbation theory \cite{PhysRev.92.609}. In a recent work \cite{PhysRevB.102.035123}, the HEG was extended to arbitrary integer dimensions. It was found a very different behavior for $D > 3$: the leading term of the correlation energy does not depend on the logarithm of $r_s$ [as in Eq.~\eqref{formula1}], but instead scales polynomially as $ c_D /r_s^{\gamma_D}$, with the exponent $\gamma_D=(D-3)/(D-1)$. In the large-$D$ limit, the value of $c_D$ was found to depend linearly with the dimension. This result was originally obtained within the random-phase approximation (RPA) by summing all the ring diagrams to infinite order. 
 
 While RPA is known to be exact in the limit of the dense gas, includes long-range interactions automatically, and is applicable to systems where finite-order many-body perturbation theories break down \cite{Ren2012}, it has well-known deficiencies at the metallic (intermediate) and low densities of the typical HEL $(1 \leq r_s \leq 6)$. Quantum Monte-Carlo (QMC) is an option for those regimes \cite{gubernatis_kawashima_werner_2016,PhysRevB.88.085121}, but there are  other high-quality approaches such as the celebrated Singwi, Tosi, Land, and Sjölander (STLS) method that provides results comparable to QMC~\cite{PhysRev.176.589}. This method attempts to tackle in an approximate manner both the exchange and electronic correlations through a local-field correction. As such, this scheme is often surprisingly accurate in the description of the full electronic density response and is commonly used to investigate qua\-si-one-dimensional \cite{Agosti}, inhomogeneneuos \cite{B904385N,Kosugi,Dharma-wardana_1981} and warm dense electron liquids \cite{Tanaka,PhysRevResearch.4.033018,DORNHEIM20181}. It has also inspired the development of new functional methods that explicitly retain the dynamical and non-local nature of electronic correlations, while properly accounting for the exchange contribution \cite{Dobson,PhysRevB.49.13267,Valenti,Yoshizawa_2009,Hasegawa}. 

The purpose of this paper is to calculate the correlation energy of the HEL for arbitrary integer dimensions (in particular for $D > 3$) with the STLS method. As expected, the value of this portion of the energy improves significantly with respect to the RPA result (that tends to over-correlate the liquid). The remainder of this paper is organized as follows.  In Section II we review the main STLS equations and rewrite them explicitly in arbitrary $D$ dimensions. As a result, we can compute the Lindhard polarizability, the structure factor, and the so-called \textit{local field correction} in the Hartree-Fock approximation, providing explicit formulae for some representative systems. In Section III we explain how the correlation energy is computed in this scheme, and discuss the fully polarized case. In Sections V and VI we present and discuss our numerical results for the exchange and correlation energies, the pair density, and its Fourier transform. We also discuss the fulfillment of the compressibility sum rule. Finally, in Section VII we present our main conclusions. Two  Appendices that give further technical details on our calculations are presented at the end of the text. 

\section{STLS Theory in $D$ dimensions}

In this section we review the main STLS equations and write them  explicitly in arbitrary $D$ dimensions. We follow the standard notation, namely, $n$ is the $D$-di\-men\-sional particle density, $\Omega$ is the volume occupied by the electronic liquid, and $q_F$ is the usual Fermi wavelength. The STLS theory departs from RPA (to include short-range correlation between electrons) by writing the two-particle density distribution $f_2(\rr,\pp, \rr',\pp', t)$ as follows: 
\begin{align}
    f_2(\rr,\pp, \rr',\pp', t) = f(\rr,\pp, t) f(\rr',\pp', t) g(\rr-\rr')\,,
    \label{eq1}
\end{align}
where $f(\rr,\pp, t)$ is the one-particle phase-space density and $g(\rr)$ is the equilibrium, static pair distribution function. Eq.~\eqref{eq1} can be seen as an ansatz that terminates the hierarchy that otherwise would write the two-particle distribution function in terms of the three-particle distribution function, and so on. This leads to the following density-density response function \cite{PhysRev.176.589}:
\begin{align}
    \chi_D(\qq, \omega) = \frac{\chi^0_D(\qq, \omega)}{1 - \Phi(\qq) [1 - G_D(\qq)] \chi^0_D(\qq, \omega)}\,.
     \label{eq2}
\end{align}
Here $\chi^0_D(\qq, \omega)$ is the Lindhard polarizability, i.e., the inhomogeneous non-interacting
density response function of an ideal Fermi gas in $D$ dimensions, $G_D(\qq)$ is the local field correction, and $\Phi(\qq)$ is the $D$-di\-men\-sional Fourier transform of the Coulomb potential:
\begin{align}
    \Phi(\qq) = \frac{(4 \pi)^{\frac{D-1}{2}} \Gamma\left( \frac{D-1}{2} \right)}{q^{D-1}}  \,,
\end{align}
where $\Gamma$ denotes the gamma function. The presence of $G_D(\qq)$ in Eq.~\eqref{eq2} is the key feature of the STLS equations that gives the ``beyond RPA'' flavor to the theory. This \textit{local field correction} is a direct result of the short-range correlation between the electrons. In arbitrary dimensions, it is given by:
\begin{align}
    G_D(\qq) = - \frac{1}{n} \int \qq' \cdot \qq \frac{q^{D-3}}{q'^{D-1}} \left[ S_D(\qq - \qq') -1\right] \frac{d^D \qq'}{(2\pi)^D} \,,
    \label{lfc}
\end{align}
with $S_D(\qq)$ being  the structure factor.
We can simplify this integral to a two-dimensional one by substituting $\qq - \qq' = \vt$ and using the fact that $S_D(\qq) = S_D(q)$ in homogeneous systems. Afterwards, we rewrite the integral using $D$-dimensional spherical coordinates, where $\qq$ is parallel to the $D^{\text{th}}$-axis, and integrate over all angles except the angle $\theta$ between $\qq$ and $\vt$ to obtain:
\begin{align}
    G_D(\qq) = - &\frac{q^{D-3}}{(2\pi)^D n}  \frac{2 \pi^{\frac{D-1}{2}}}{\Gamma \left( \frac{D-1}{2} \right)} \int_0^{\infty}   \int_0^{\pi} [S_D(t) -1] \times \notag \\ &\frac{[q^2 t^{D-1}  -qt^D \cos \theta] (\sin \theta)^{D-2}}{(q^2 + t^2 - 2qt\cos\theta)^{\tfrac{D-1}{2}}} d\theta dt \,.
    \label{eq3}
\end{align}
Note in passing that in the $2D$ case we must use polar coordinates instead of the spherical coordinates and obtain:
\begin{align}
    G_{2}(\qq) = -\frac{1}{(2\pi)^2 n} &\int_0^{\infty} \int_0^{2\pi} [S_{2}(t)-1] \times \notag \\ &\frac{qt - t^2 \cos\theta}{(q^2 + t^2 - 2qt \cos \theta)^{1/2}} d\theta dt \,.
\end{align}

These equations, together with the equation for the dielectric function, $1/\epsilon_D(\qq,\omega) = 1+ \Phi(\qq)\chi_D(\qq,\omega)$, lead to an equation for the dielectric function within the STLS theory, namely:
\begin{align}
    \epsilon_D(\qq,\omega) = 1 - \frac{ \Phi(\qq)\chi^0_D(\qq,\omega)}{1 + G_D(\qq)\Phi(\qq)\chi^0_D(\qq,\omega)} \,.
    \label{eq4}
\end{align}

Finally, the relation between the structure factor and the dielectric function $\epsilon_D(\qq, \omega)$ (see, for instance, \cite{pines1999elementary}) can be easily generalized for arbitrary dimensions:
\begin{align}
    S_D(\qq)= - \frac{1}{\pi n \Phi(\qq)} \int_0^\infty {\rm Im} \left(\frac{1}{\epsilon_D(\qq,\omega)}\right)d\omega \,.
    \label{eq5a}
\end{align}
We can improve the readability of this equation by separating the contributions of the single-particle and plasmon excitations~\cite{pinesnozieres}:
\begin{align}
    S_D(\qq) &= - \frac{1}{\pi n \Phi(\qq)} \int_0^{ qv_F + \frac{q^2}{2}} {\textrm {Im}} \left(\frac{1}{\epsilon_D(\qq,\omega)}\right)d\omega \notag \\ + &\frac{1}{\Phi(\qq) n} \left(\frac{\partial \textrm{Re} \epsilon_D(\qq, \omega)}{\partial \omega} \right)^{-1} \delta(\omega - \omega_p(q)) \,,
    \label{eq5b}
\end{align}
where $v_F$ is the Fermi velocity and $\omega_p(q)$ is the plasmon dispersion \cite{Bernadotte2013}.
Equation~\ref{eq5b}, Eq.~\ref{eq3} and Eq.~\ref{eq4} form the core of the STLS set of equations. In this framework, they can be evaluated \emph{self-consistently}. Notice, indeed, that $G_D(\qq)$ and $S_D(\qq)$ can be written symbolically as $G_D = \mathcal{F}_1[S_D]$ and $S_D = \mathcal{F}_2[G_D]$, indicating the existence of the mutual functional relations introduced above \cite{Yoshizawa_2009}.

We should reiterate here that the crucial aspect of the STLS method is the appearance of the density-density pair distribution $g(\rr-\rr')$ in the decoupling of the equation of motion \eqref{eq1}. Despite the crudeness of such a factorization scheme, the formalism gives excellent
correlation energies for both 3D and 2D homogeneous electron gases. Unfortunately, the factorization also results in a number of well-known shortcomings, including negative values of $g(\rr-\rr')$ for sufficiently large values of $r_s$, and the failure to satisfy the compressibility sum rule (i.e., at long wavelengths the exact screened density response is determined by the isothermal compressibility). The first problem, the un-physical behavior of the pair distribution, is counter-balanced by the fact that the exchange-correlation energy is in reality an integral over $g(r)$: it turns out that its value is still quite reasonable in the metallic range as it benefits from an error cancellation \cite{DORNHEIM20181}. As we will see below, this result, known for the case of 2D and 3D, does also hold for larger dimensions. To tackle the second problem, Vashista and Singwi \cite{PhysRevB.6.875} provided a correction to the STLS method that established the correct compressibility sum rule in the metallic-density regime. This is however a partial solution as both the original STLS method and the Vashista-Singwi extension under-estimate the exchange energy, as pointed out by Sham \cite{PhysRevB.7.4357}.

\subsection{Real part of the Lindhard function for $D = 5, 7$}

 The expressions for $\chi^0_{2}(\qq, \omega)$ and  $\chi^0_{3}(\qq, \omega)$ are widely known since long ago (see for instance \cite{giuliani_vignale_2005,Haug}), but higher dimensional expressions are missing in the literature. In Appendix \ref{app1} we detail their calculation for higher dimensional settings by performing a linear perturbation from equilibrium. We sketch here the main results.
\begin{widetext}
The real part of $\chi^0_D(\qq, \omega)$ can be written explicitly as:
\begin{align}
    \mathrm{Re} \chi^0_D (\qq, \omega) &= \frac{2}{(2 \pi)^D} \mathcal{P} \int \frac{\Theta(q_F - |\pp - \tfrac{1}{2}\qq|) \Theta(|\pp + \tfrac{1}{2}\qq|- q_F)}{ \omega - \pp \cdot \qq} d^D \pp \notag \\ & \quad - \frac{2}{(2 \pi)^D} \mathcal{P} \int \frac{\Theta(|\pp - \tfrac{1}{2}\qq| - q_F) \Theta(q_F - |\pp + \tfrac{1}{2} \qq|) }{ \omega - \pp \cdot \qq} d^D\pp \,,
\end{align}
where $\mathcal{P}$ denotes the principal value. By evaluating these integrals, it is possible to obtain analytical expressions for specific cases. For instance, for $D = 5$ and $D=7$ one gets:
\begin{align}
    \mathrm{Re} \chi^0_{5} (\qq, \omega) = \frac{ q_F^3}{8 \pi^3} \Bigg\{&\frac{1}{96 \qn^5} \Bigg[\left(\frac{3}{2}\left(\qn^2 -2\wn\right)^4 + 24\qn^4 -12\qn^2(\qn^2-2\wn)^2 \right) \ln \left| \frac{2\qn - \qn^2 + 2\wn}{2\qn + \qn^2 - 2\wn} \right| \notag \\
    + &\left(\frac{3}{2}\left(\qn^2 +2\wn \right)^4 + 24\qn^4 -12\qn^2(\qn^2+2\wn)^2 \right) \ln \left| \frac{2\qn - \qn^2 - 2\wn}{2\qn + \qn^2 + 2\wn} \right| +12\qn^7 -16\qn^5 + 144\qn^3 \wn^2  \Bigg] - \frac{2}{3}\Bigg\} 
    \label{LPD5}
\end{align}
and
\begin{align}
    \mathrm{Re} \chi^0_{7} (\qq, \omega) = \frac{ q_F^5}{368640 \pi^4 \qn^5}\Bigg\{ &\left[60(16\qn^4 + 3(\qn^2 + 2\wn)^4 - 12(\qn^3 + 2\qn\wn)^2) - \frac{15(\qn^2 + 2\wn)^6}{\qn^2} \right] \ln \left|\frac{2\qn - \qn^2 - 2\wn}{2\qn + \qn^2 + 2\wn} \right| \notag \\ + &\left[60(16\qn^4 + 3(\qn^2 - 2\wn)^4 - 12(\qn^3 - 2\qn\wn)^2) - \frac{15(\qn^2 - 2\wn)^6}{\qn^2} \right] \ln \left|\frac{2\qn - \qn^2 + 2\wn}{2\qn + \qn^2 - 2\wn} \right| \notag \\ - &4224\qn^5 + 1280\qn^3(\qn^4 + 12\wn^2) - 120\qn(\qn^8 + 40 \qn^4 \wn^2 + 80\wn^4) \Bigg\}\,.
    \label{LPD7}
\end{align}
A similar calculation gives a closed expression for the imaginary part:
\begin{align}
\label{impart}
    \mathrm{Im} \chi^0_D(\qq, \omega) =
    \begin{cases}
         h(D) \dfrac{1}{\qn} \left[ \left(1- \nu_{-}^2 \right)^{\frac{D-1}{2}} - \left( 1-  \nu_{+}^2\right)^{\frac{D-1}{2}}\right] & \wn < \left|\qn - \frac{\qn^2}{2} \right| \text{ and } \qn < 2\\
         0 & \wn < \left|\qn - \frac{\qn^2}{2} \right| \text{ and } \qn > 2 \\
         h(D) \dfrac{1}{\qn} \left[1- \nu_{-}^2 \right]^{\tfrac{D-1}{2}} & \left|\qn - \frac{\qn^2}{2}\right| \leq \wn \leq \qn + \frac{\qn^2}{2} \\
         0 & \wn > \qn + \frac{\qn^2}{2}
    \end{cases}
\end{align}
where $\qn = q/q_F$, $\wn = \omega/q_F^2$, $h(D) = q_F^{D-2}\left[2^{D-2} (D-1) \pi^{\tfrac{D-1}{2}} \Gamma \left( \tfrac{D-1}{2} \right)\right]^{-1}$ and $\nu_{\pm} = \wn/\qn \pm \qn/2$.
\end{widetext}

\subsection{The Hartree-Fock approximation for the first iteration}
\label{sectionHF}

For the first iteration, the original STLS theory uses the structure factor obtained  from the Hartree-Fock calculation~\cite{PhysRev.176.589}. We take the same approach here. Formally, the generalized $D$-dimensional structure factor is straightforward and reads:
\begin{align*}
    S^{\textrm{HF}}_D(\qq) = 1 - \frac{2}{(2 \pi)^D \, n} \int_{k,k' \leq q_F}  \delta(\qq - \kk' + \kk) \, d^D \kk' d^D \kk \, .
\end{align*}
Substituting this formal result into Eq.~\eqref{lfc} gives the general $D$-dimensional expression for the local field correction of STLS within the Hartree-Fock approximation, namely,
\begin{align}
    G_D^{\textrm{HF}}(\qq) &= \frac{2 q^{D-3}}{(2 \pi)^{2D} n^2} \int_{k, k' \leq q_F} \frac{\qq \cdot (\qq + \kk - \kk')}{|\qq + \kk - \kk'|^{D-1}} \, d^D \kk' d^D \kk\,.
    \label{eqGHF}
\end{align}
This integral can be evaluated by making use of the extracule and intracule substitutions: $\vs = (\kk + \kk')/2 \text{ and } \vt = \kk - \kk'$. Then, because the integrand only depends on $\qq$ and $\vt$, one can perform the integration over $\vs$. This eventually gives the relevant integration region as the intersection of two hyperspheres, as discussed in detail in Appendix \ref{app2}, where we obtain the following expression:
\begin{multline}
    G_D^{\textrm{HF}}(\qq) = \frac{q^{D-3}}{q_F^D \sqrt{\pi}}\frac{\Gamma \left( \frac{D}{2} + 1\right)}{\Gamma \left( \frac{D-1}{2}\right)} \int_0^{2q_F} \!\!\int_0^{\pi} \textrm{I}_{1-\frac{t^2}{4q_F^2}}\!\! \left(\frac{D+1}{2}, \frac{1}{2} \right) \\ 
    \times \frac{(q^2 + qt \cos \theta) t^{D-1} \sin^{D-2}\theta}{(q^2 + t^2 + 2qt \cos \theta)^{\frac{D-1}{2}}} d\theta dt \,,
    \label{eqGHF2}
\end{multline}
with $I_x(a,b)$ denoting the regularized incomplete beta function. The numerical evaluation of the local field correction as expressed in Eq.~\eqref{eqGHF2} is now a much simpler task. In fact, we present the lengthy analytical expression for $G_5^{\textrm{HF}}(\qq)$ in Appendix \ref{app2}.

\section{Energy Contributions}

The calculation of the kinetic energy of the $D$-di\-men\-sional HEG is as straightforward as it is for $D = 3$. It is given by:
\begin{align}
    E_{\textrm{kin}} = \frac{\alpha_D^2 D}{2(D+2)} \frac{\Upsilon_2 (\xi)}{r_s^2} \,,
    \label{ekin}
\end{align}
where $\alpha_D = 2^{(D-1)/D} \Gamma(D/2 + 1)^{2/D}$,
$$\Upsilon_n(\xi) = \tfrac{1}{2}\left[(1 + \xi)^{(D+n)/D} + (1-\xi)^{(D+n)/D} \right]$$ is the spin-scaling function, and $\xi$ is the usual spin-po\-la\-ri\-zation. The interaction energy per particle is given by the following expression:
\begin{align}
    E_{\textrm{int}} = \frac{1}{\Omega} \sum_{\qq \neq 0} \frac{\Phi(\qq)}{2} (S(\qq) - 1)\,.
    \label{eq7}
\end{align}
If we introduce the function $\gamma(r_s) = -\tfrac{1}{2q_F} \int_0^{\infty} [S(q) - 1] dq$ we can rewrite the energy of the unpolarized electron gas in Eq.~\eqref{eq7} as:
\begin{align*}
    E_{\textrm{int}} = - \frac{2^{\tfrac{D-3}{D}} D^{\tfrac{2}{D}}}{\sqrt{\pi}} \Gamma\left( \frac{D-1}{2} \right) \Gamma\left( \frac{D}{2} \right)^{\tfrac{2-D}{D}} \frac{\gamma(r_s)}{r_s}\,.
\end{align*}
By using the adiabatic-connection formula, the total interaction energy per particle can be computed by using the formula \cite{PhysRevLett.1.443}:
$\int_0^1  \lambda E_{\textrm {int}}(\lambda r_s) d\lambda$, where $\lambda$ is a coupling constant that represents the strength of the interaction. Therefore, the full ground-state energy of the unpolarized electron gas in the STLS theory can then be written as:
\begin{align}
    E_0 &= \frac{\alpha_D^2 D}{2(D+2)} \frac{\Upsilon_2 (0)}{r_s^2} - E(D) \frac{1}{r_s^2} \int_0^{r_s} \gamma(r_s')dr_s' \,,
\end{align}
where $E(D)$ is defined as:
\begin{align}
    E(D) = \frac{2^{\tfrac{D-3}{D}} D^{\tfrac{2}{D}}}{\sqrt{\pi}} \Gamma\left( \frac{D-1}{2} \right) \Gamma\left( \frac{D}{2} \right)^{\tfrac{2-D}{D}}\,.
    \label{ed}
\end{align}

To obtain the correlation energy we need to make use of the Hartree-Fock energy. While the kinetic energy is already calculated, the calculation of the exchange energy is not trivial. This was already calculated by two of us in Ref.~\cite{PhysRevB.102.035123}:
\begin{align}
    E_{\textrm{HF}} = \frac{\alpha_D^2 D}{2(D+2)} \frac{\Upsilon_2 (\xi)}{r_s^2}  - \frac{2\alpha_D D}{\pi (D^2-1)} \frac{\Upsilon_1(\xi)}{r_s}\,.
\end{align}
Since the correlation energy is defined as the difference between the ground-state energy and the ground-state Hartree-Fock energy, we get for the $D$-dimensional gas the following compact formula:
\begin{align}
    E_{\textrm{corr}} = \frac{1}{r_s^2} \int_0^{r_s} \left[- E(D)\gamma(r_s') + \frac{2\alpha_D D \Upsilon_1(0)}{\pi (D^2-1)} \right]dr_s'\,.
    \label{eq8}
\end{align}
This is the formula we are going to use for the calculation of the correlation energy. 

\section{The fully polarized case}
The extension of the above equations for the fully polarized case is quite straightforward (for $D=3$ see \cite{PhysRevB.79.115304,PhysRevB.35.6683}). The relation between the local field correction and the structure factor is given by:
\begin{align}
    G^{\uparrow\uparrow}_D(q) =  -\frac{1}{n} \int \frac{(\qq' \cdot \qq)q^{D-3}}{q'^{D-1}}  [S^{\uparrow\uparrow}_D(\qq -\qq') -1] \frac{d^D\qq'}{(2\pi)^D} \, ,
\end{align}
where $S^{\uparrow\uparrow}_D(\qq -\qq')$ is the spin-resolved structure factor. The corresponding density-density response function is given by:
\begin{align}
    \chi^{\uparrow\uparrow}_D(\qq, \omega) = \frac{\chi^{0,\uparrow \uparrow}_D(\qq, \omega)}{1 - \Phi(\qq)[1 - G^{\uparrow \uparrow}_D(\qq)]\chi^{0,\uparrow \uparrow}_D(\qq, \omega)} \,.
\end{align}
where $\chi^{0,\uparrow \uparrow}_D(\qq, \omega)$ is the polarizability of the fully polarized non-interacting HEG that can be easily related with the spinless quantities $\chi^{0}_D(\qq, \omega)$ by a factor of $1/2$ \cite{giuliani_vignale_2005}. In addition, the Fermi wavelength should be re-scaled $q_{F,\uparrow} =2^{1/D}q_{F}$. Finally,  the fluctuation-dissipation theorem leads to:
\begin{align}
    S^{\uparrow\uparrow}_D(\qq) = - \frac{1}{\pi n} \int_0^{\infty} \mathrm{Im} \chi^{ \uparrow\uparrow}_D (\qq, \omega) d \omega\,.
\end{align}

As a result, in the fully polarized case, the correlation energy is given by:
\begin{align}
    E_{\textrm{corr}} = \frac{1}{r_s^2} \int_0^{r_s} \left[- F(D)\gamma(r_s') + \frac{2\alpha_D D \Upsilon_1(1)}{\pi (D^2-1)} \right]dr_s'\,.
\end{align}
where $F(D)$ is a simple re-scaling of $E(D)$ in Eq.~\eqref{ed}:
\begin{align}
    F(D) =2^{1/D} E(D)\,.
\end{align}

\section{Numerical Results}

Starting from the expression for the local field correction in the Hartree-Fock approximation $ G_D^{\textrm{HF}}(\qq)$ presented in section \ref{sectionHF} we calculated $S_D(\qq)$ using Eq.~\eqref{eq5b}. In this case $\omega(p)$ is obtained from the zero of the dielectric function $\epsilon(\qq, \omega)$, which in turn is given by Eq.~\eqref{eq4}. We then started the entire self-consistent cycle of the STLS equations. To improve the convergence of the iterative procedure we applied the following linear mixing:
\begin{equation}
  \bar{G}_D^i(\qq) = G_D^{i-1}(\qq) + [G_D^i(\qq) - G_D^{i-1}(\qq)]/a
\,,
\end{equation}
where $i$ is the iteration number and $a$ takes values between 1 and 3.5 (for the $5D$ and the $7D$ cases we used $a=1.5$). At each iteration, we calculated the quantity $\gamma(r_s)$ with the new structure factor $S_D(\qq)$. After 10 iterations we obtained convergence in $\gamma(r_s)$ within 0.1\%. The value of $\gamma(r_s)$ at the end of the self-consistent calculation is then used to calculate the correlation energy using Eq.~\eqref{eq8}. 

\begin{figure}[!t]
    \includegraphics[width =\linewidth]{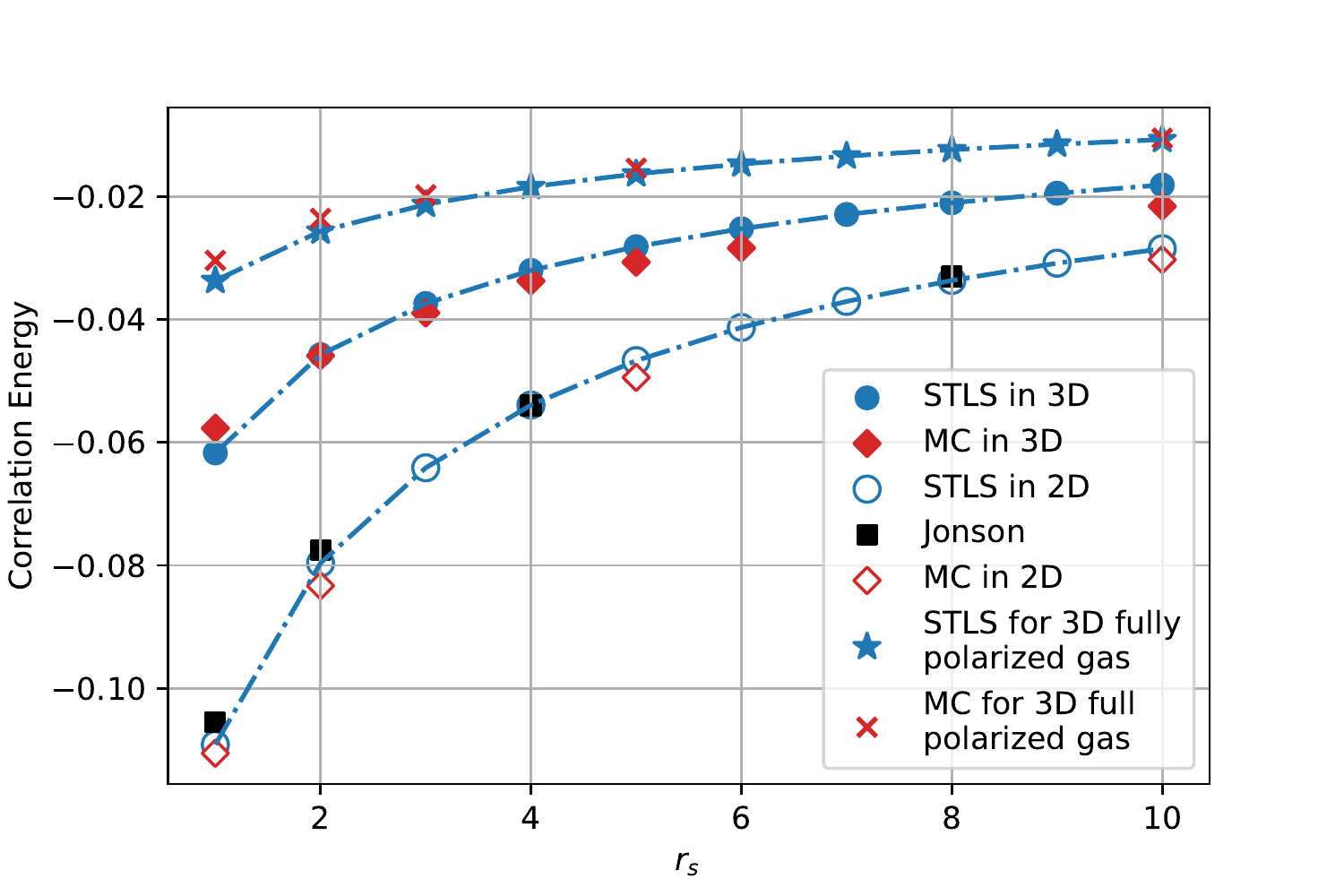}
    \caption{Correlation energy for HEG in $3D$ and $2D$ expressed as a function of $r_s$. Our results for STLS in the paramagnetic case are compared with the results of Jonson \cite{MJohnson} and the results of QMC from Refs.~\cite{PhysRevB.18.3126} and \cite{PhysRevA.6.393}. In the ferromagnetic case our results for STLS are compared with the results of QMC from Ref.\cite{PhysRevB.88.085121}.}
    \label{fig2D_3D}
\end{figure}

We first benchmarked our implementation for $2D$ and $3D$. The results are presented in Fig.~\ref{fig2D_3D}, where it can be seen that we recovered the original STLS results and obtained the celebrated agreement of STLS with the Quantum Monte-Carlo results \cite{PhysRevB.18.3126,PhysRevA.6.393}. In the $2D$ case we also obtained similar results to the ones obtained previously by Jonson \cite{MJohnson}. The decrease of the magnitude of the correlation energy in the paramagnetic case is consistent with the Quantum Monte-Carlo results \cite{PhysRevB.88.085121}.

\begin{figure}[!t]
    \begin{subfigure}{\linewidth}
        \centering
        \includegraphics[width =\linewidth]{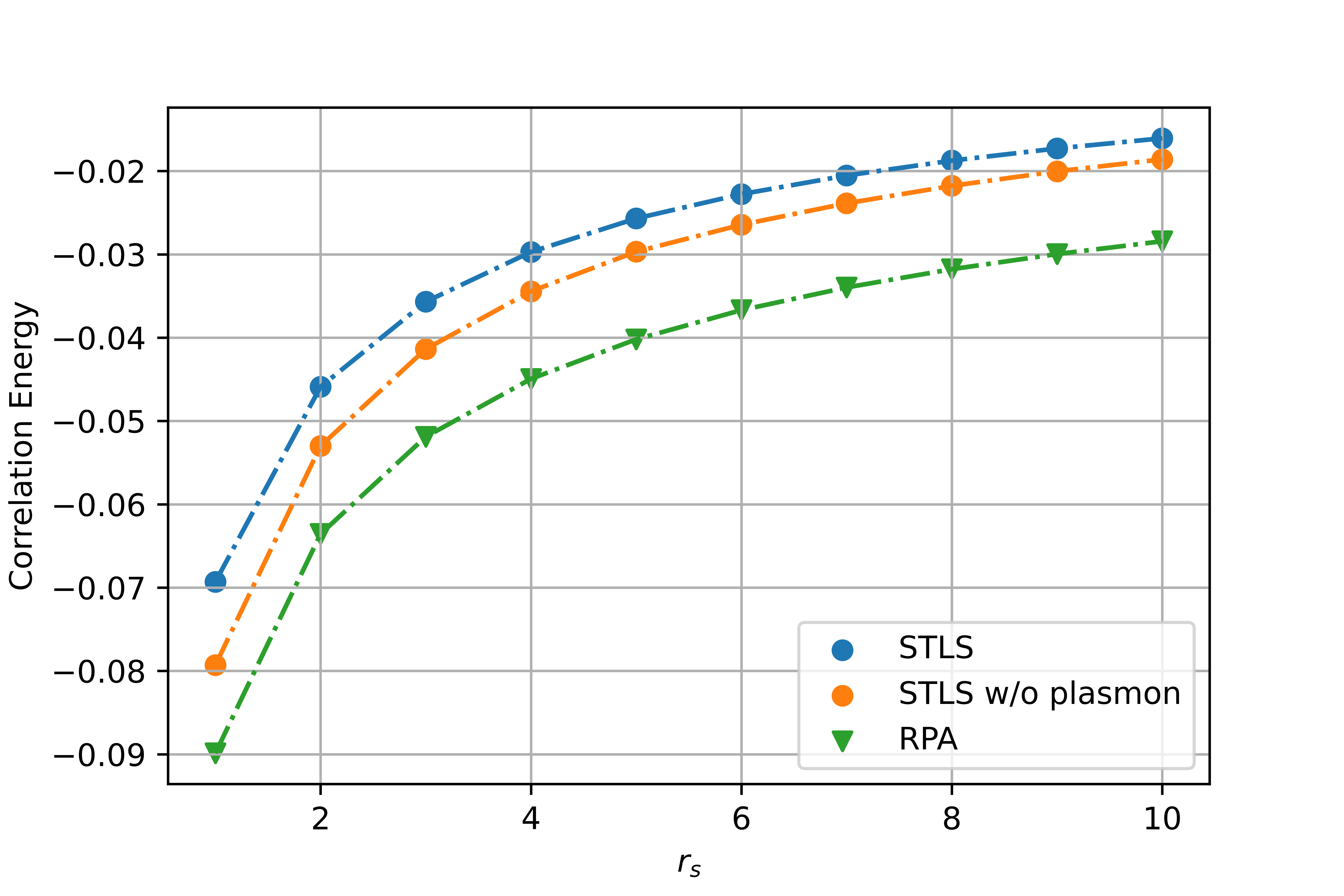}
        \caption{HEG in 5D}
        \label{fig5D}
    \end{subfigure}
    \begin{subfigure}{\linewidth}
        \centering
        \includegraphics[width =\linewidth]{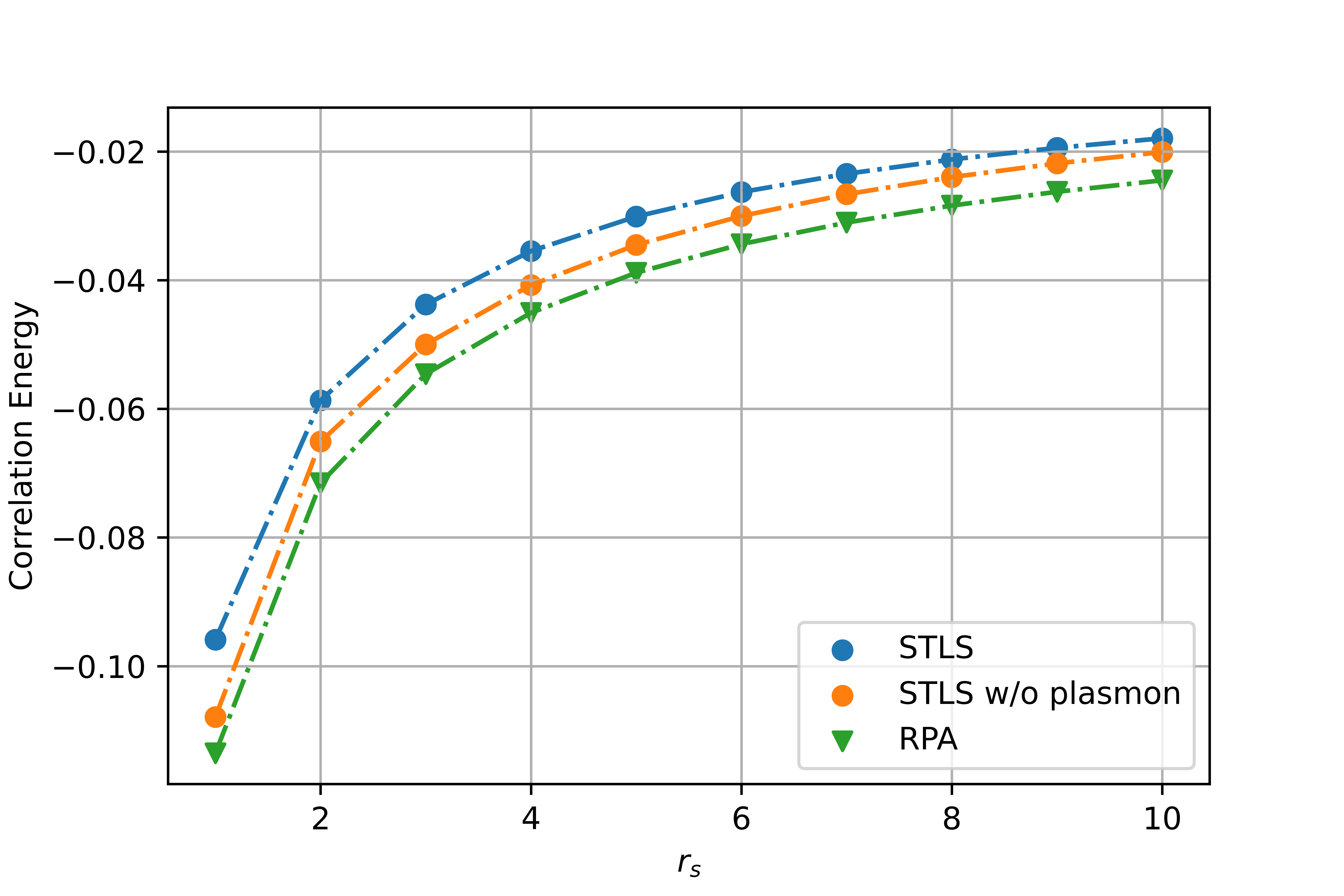}
        \caption{HEG in 7D}
        \label{fig7D}
    \end{subfigure}
    \caption{Correlation energy of the unpolarized HEG in $5D$ and $7D$ expressed as a function of $r_s$. Our results for STLS with and without the plasmon contribution are compared with RPA results from Ref.~\cite{PhysRevB.102.035123}.}
\end{figure}

\begin{figure}[!t]
    \begin{subfigure}{\linewidth}
        \centering
        \includegraphics[width =\linewidth]{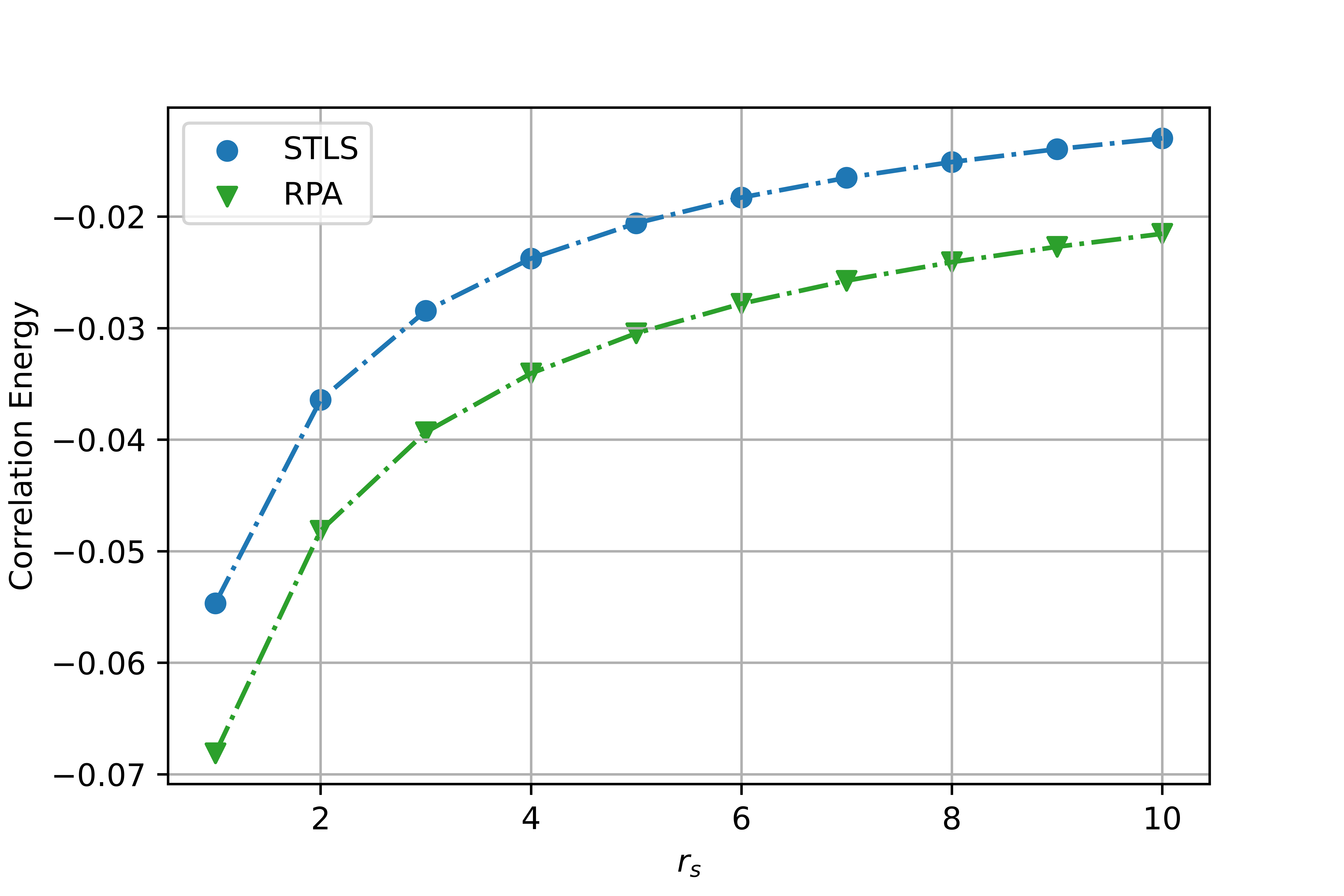}
        \caption{HEG in 5D}
        \label{fig5Dp}
    \end{subfigure}
    \begin{subfigure}{\linewidth}
        \centering
        \includegraphics[width =\linewidth]{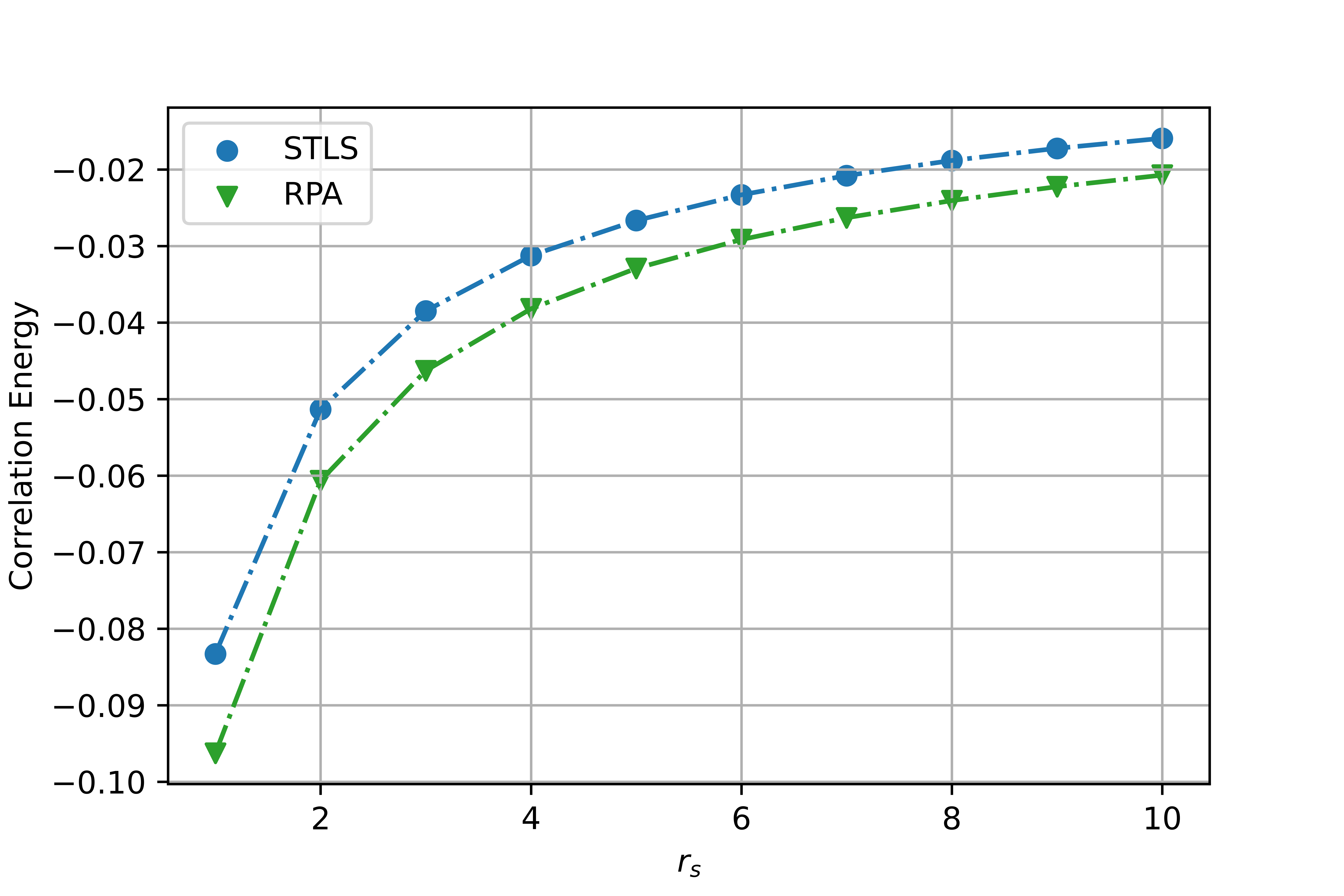}
        \caption{HEG in 7D}
        \label{fig7Dp}
    \end{subfigure}
    \caption{Correlation energy of the fully polarized HEG in $5D$ and $7D$ expressed as a function of $r_s$. Our results for STLS are compared with the RPA results from Ref.~\cite{PhysRevB.102.035123}.}
    \label{fully_polar}
\end{figure}

Our new results for the HEG in 5D and 7D (for which we were able to compute analytically the Lindhard polarizability and the local field correction) in the paramagnetic and the ferromagnetic case are presented in Fig.~\ref{fig5D}, \ref{fig7D}, \ref{fig5Dp} and \ref{fig7Dp}, respectively. Those are compared with the RPA results previously obtained in Ref.~\cite{PhysRevB.102.035123} \footnote{We corrected a small error in the numerical evaluation of the coefficients presented in Table I of Ref.~\cite{PhysRevB.102.035123} that however does not affect the main conclusions of that paper.}. It is a general result of our implementation that the STLS correlation energy decreases (in absolute value) in comparison with the one from RPA, which confirms that this latter approximation tends to over-correlate the gas, even in dimensions higher than 3. Yet, for high dimensions the general error of RPA is smaller, both in the ferromagnetic and the paramagnetic cases. The correlation energy is smaller in magnitude in the ferromagnetic case in comparison with the paramagnetic one in all the dimensions studied, the same behaviour found with the RPA. Furthermore, since the structure factor of STLS can be easily separated into the pair and the plasmonic contributions \cite{pines1999elementary,Bernadotte2013}, we investigated the plasmon contribution separately by calculating the correlation energy with and without it. We conclude that the plasmonic correction is more relevant for intermediate densities at all dimensions: this term indeed improves the correlation energy on average by 30.9\% in the 5D case and 49.2\% in the 7D case.    

\begin{figure}[!t]
    \centering
    \includegraphics[width = \linewidth]{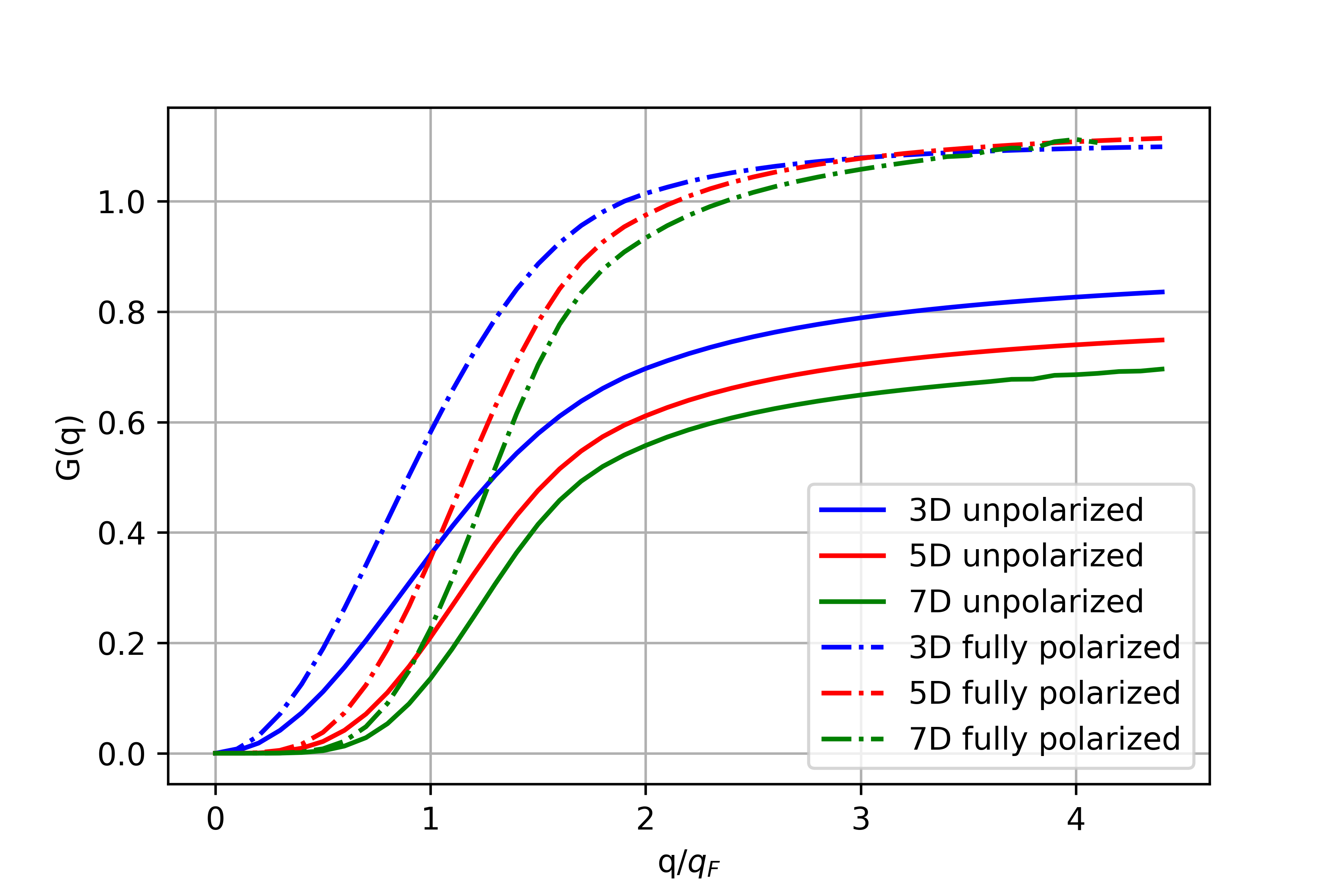}
    \caption{The local field correction $G(\qq)$ for $r_s = 2$ in the $3D$, $5D$ and $7D$ case for the para- and ferromagnetic HEG.}
    \label{LCF}
\end{figure}

For the sake of completeness, we also present in Fig.~\ref{LCF} the values for the local field correction $G(\qq)$ we obtained for both paramagnetic and ferromagnetic gases for $D=3, 5, 7$ for a selected value of the density (i.e., $r_s = 2$). While the magnitude of the local correction is larger for the fully polarized gas, the global value diminishes for large dimensions in the whole domain of $q$. 

\begin{figure}[b]
  \centering
  \includegraphics[width=0.8\linewidth]{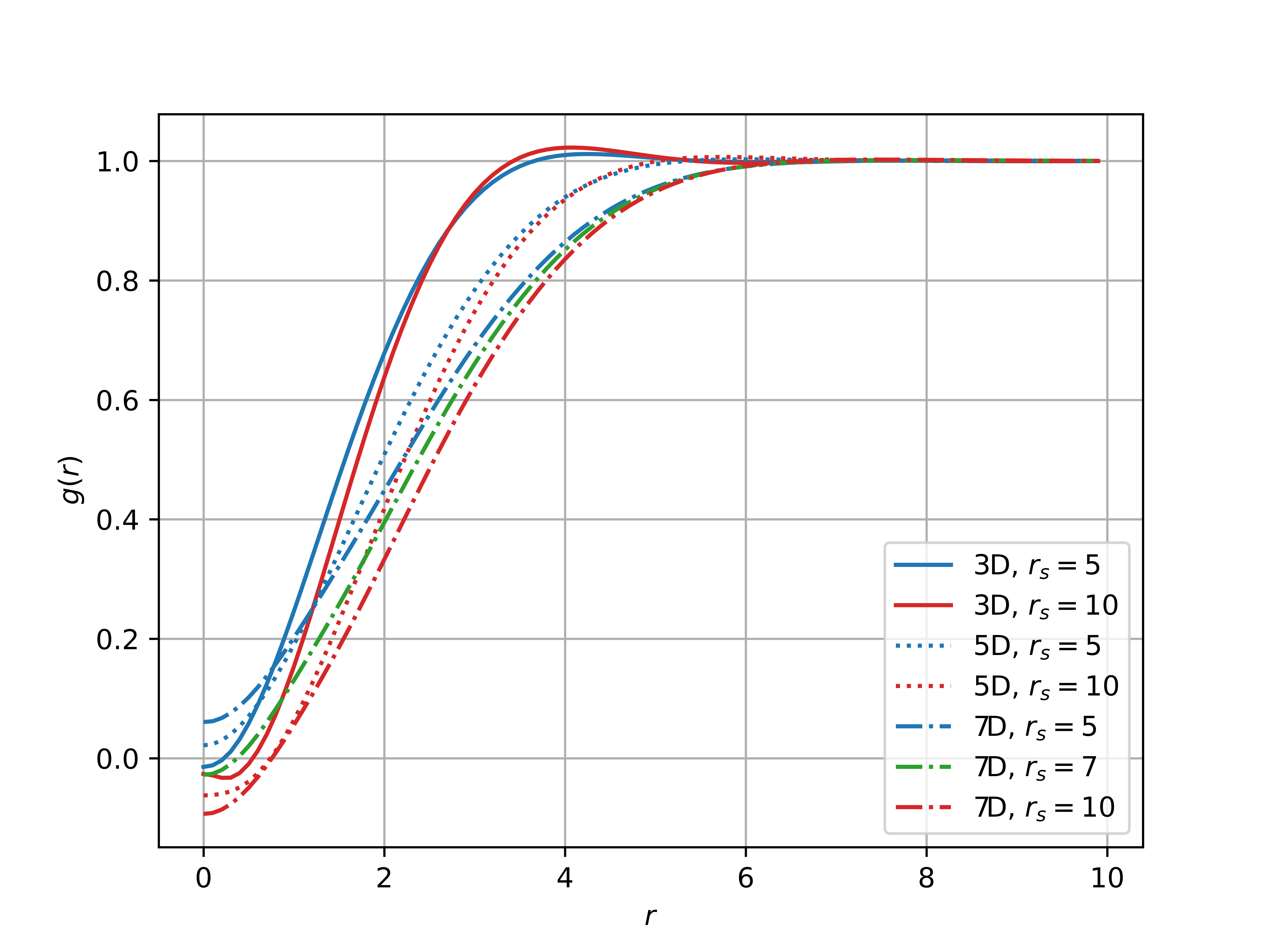}
\caption{The density-density pair distribution $g(r)$ of the HEL at $r_s = 5$ and $10$ for dimensions $D=3,5$ and 7.}
\label{fig:5}
\end{figure}

\begin{figure}[b!]
  \centering
  \includegraphics[width=0.8\linewidth]{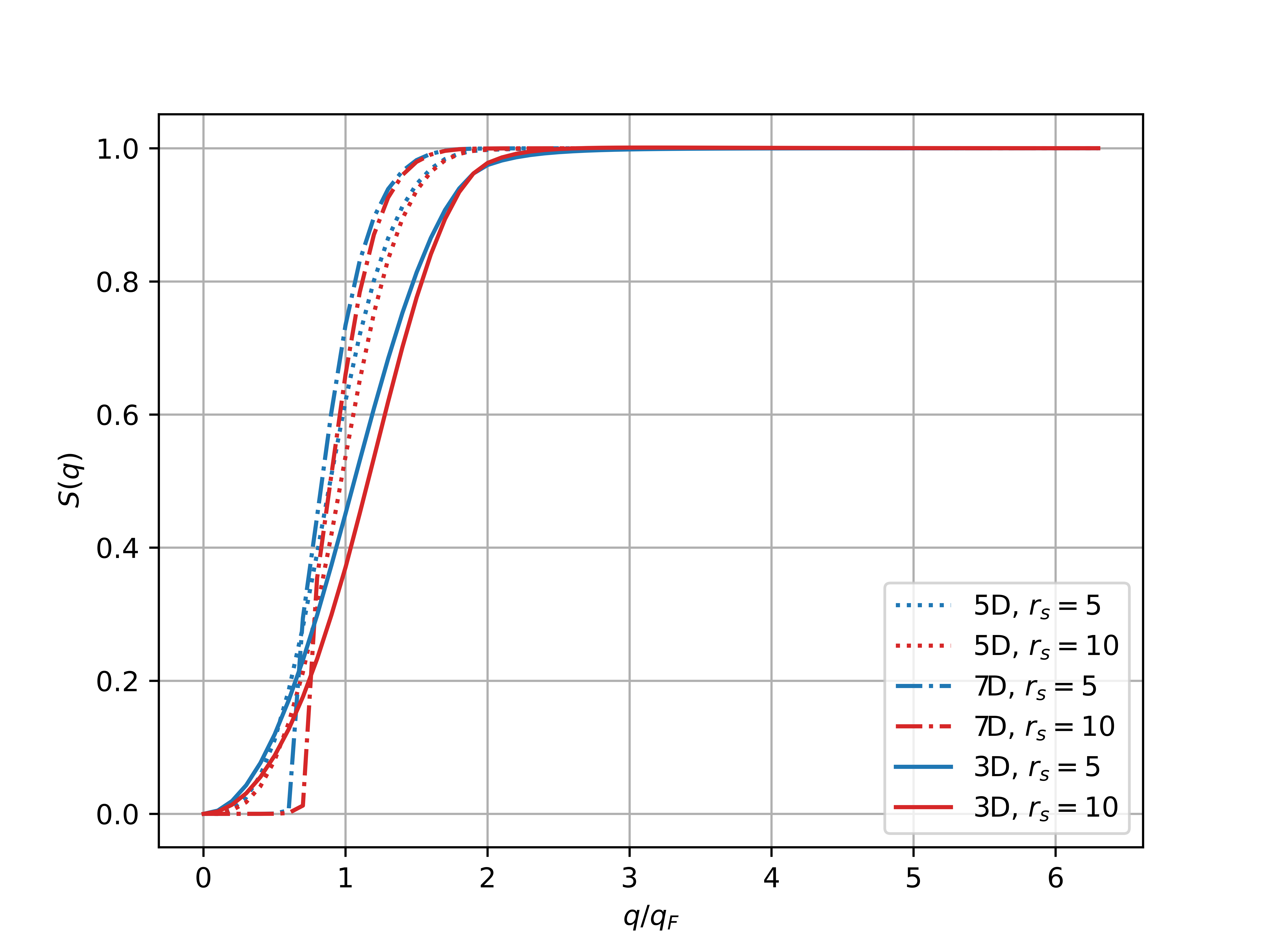}
\caption{The structure factor $S(q)$ at $r_s = 5$ and $10$ for dimensions $D=3,5$ and 7.}
\label{fig:6}
\end{figure}

Finally, to study the quality of the density-density pair distribution in Figs.~\ref{fig:5} and \ref{fig:6} we plot $g(r)$ and its Fourier transform $S(q)$ at the metallic density $r_s = 5$ and the large radius $r_s = 10$ for dimensions $D=3,5$, and $7$. The behavior of $g(r)$ for the metallic densities (i.e., $r_s = 5$) is physical in the sense that it takes positive values for all the analyzed dimensions, whereas for $r_s=10$ an un-physical, though small, negative value can be observed for small distances. This is in agreement with the 2D and 3D cases, discussed above; this also shows that the range of validity of the STLS scheme is limited to the metallic-density regime, even at unconventional dimensions.

\section{Sum rules}
\label{CSR}

As already mentioned, one of the known shortcomings of the STLS method is the failure to satisfy the compressibility sum rule (CSR) which is an exact property of the HEL: In the long wavelength limit, the static response and dielectric functions are related to the compressibility of the electronic system. Indeed, for arbitrary dimensions the following expression holds (see Appendix \ref{appCSR}):
\begin{align}
    \lim_{q\rightarrow 0 } \epsilon_D(\qq,0) = 1 + \left(\frac{q_{\rm TF}}{q}\right)^{D-1} \frac{\kappa}{\kappa_f}\,,
    \label{compren}
\end{align}
where $\kappa/\kappa_f$ is the ratio of the compressibility of the
 interacting-free $D$-dimensional electron liquid, and  $q_{\rm TF}$ is the $D$-dimensional Thomas-Fer\-mi wave vector. For the free electron liquid, the corresponding free compressibility is $\kappa_f = D r_s^2 /n\alpha_D^2$. 
 Also by taking the limit $q \rightarrow 0$ at $\omega = 0$ in Eq.~\eqref{eq4} and comparing the result with Eq.~\eqref{compren} (see also Appendix \ref{appCSR}) one obtains:
 \begin{align}
    \frac{\kappa_f}{\kappa} = 1 -  \gamma_D \left(\frac{q_{\mathrm{TF}}}{q_F} \right)^{D-1} \coloneqq \mathcal{K}_D \,,
    \label{compren2}
\end{align}
where $\gamma_D = -\frac{1}{2q_{\mathrm{F}}} \int^\infty_0 [S_D(q)-1]dq$.

 Another way to calculate the same com\-pre\-ssibility ratio in arbitrary dimensions is to use (a) the known thermody\-na\-mic formula $\kappa^{-1}  = n (\partial P /\partial n)$, where $P$ denotes the pressure, and (b) the expression of the energy for the free liquid \eqref{ekin}. Our result is 
\begin{align}
\frac{\kappa_f}{\kappa} = \frac{r_s^4}{\alpha^2_D D}\left[\frac{(1-D)}{r_s}\epsilon'_D(r_s)+\epsilon''_D(r_s)\right] \coloneqq \bar{\mathcal{K}}_D \,.
\label{termo}
\end{align}

\renewcommand{\arraystretch}{1.6}
\begin{table}[b!]
    \centering
    \begin{tabular}{c|cc|cc|cc}
          & \multicolumn{2}{c|}{{$3D$}} &\multicolumn{2}{c|}{$5D$} & \multicolumn{2}{c}{$7D$}  \\
         & {$\mathcal{K}_3$} & $\bar{\mathcal{K}}_3$ & {$\mathcal{K}_5$} & $\bar{\mathcal{K}}_5$ & {$\mathcal{K}_7$} & $\bar{\mathcal{K}_7}$ \\ \hline
         $r_s = 2$ &  0.35 & 0.64 &  0.75 & 0.88 & 0.85 & 0.93 \\  
         $r_s = 4$ & -0.39 & 0.25 &  0.46 & 0.73 & 0.69 & 0.85 \\
         $r_s = 6$ & -1.18 & -0.16 &  0.15 & 0.58 & 0.52 & 0.77
    \end{tabular}
    \caption{Compressibility ratios $\mathcal{K}_D$ \eqref{compren2} and $\bar{\mathcal{K}}_D$ \eqref{termo} predicted by the STLS scheme in $3D$, $5D$ and $7D$.}
    \label{tab_compress}
\end{table}

In practice, Eq.~\eqref{compren} implies that the compressi\-bi\-li\-ty ratio $\kappa/\kappa_f$ derived from the thermodynamic expression in Eq.~\eqref{termo} should be equal to the same ratio derived from the $\lim_{q\rightarrow 0 }  \epsilon_D(\qq,0)$ of Eq.~\eqref{eq4} with  $G_D(\qq)$ as given in Eq.~\eqref{eq3}. Yet, calculating the compressibility from an approximate expression of the dielectric function will ge\-ne\-rally result in a value that differs from the one obtained as the derivative of the pressure. This is the case of STLS theory, indeed. As a matter of fact, there are few theoretical settings where they coincide \cite{PhysRevB.88.115123}.  

The numerical results of $\kappa_f/\kappa$ in $3D$, $5D$ and $7D$ calculated both as $\mathcal{K}_D$ \eqref{compren2} and $\bar{\mathcal{K}}_D$ \eqref{termo} for some $r_s$ are shown in Table \ref{tab_compress}. In general, for all dimensions $\mathcal{K}_D \neq \bar{\mathcal{K}}_D$. For 3D we obtained the results of the original STLS paper \cite{PhysRev.176.589}. Notice that the STLS compressibility ratio $\mathcal{K}_3$ becomes already negative for $r_s > 3$. Yet, interestingly, the larger the underlying  dimension the better the performance of the CSR within the original STLS framework.  For instance, for $r_s = 4$, $\mathcal{K}_D/\bar{\mathcal{K}}_D = -1.56 , 0.63$, and $0.81$  for $3D$, $5D$ and $7D$, respectively.


There are other well-known sum rules that can be discussed in the context of STLS  \cite{PhysRevB.7.4300}. For instance, in the limit of large frequencies, one can show that $\epsilon_D(\qq,\omega) = 1- [\omega'_D(\qq)/\omega]^2$,
where $[\omega'_D(\qq)]^2 = \Phi(\qq) n q^2$ is a frequency that in the case of $D=3$ is equal to $\omega^2_p = 4\pi n/m$, the plasma frequency of the liquid. Moreover, by using however the results of Appendix \ref{appCSR} and the frequency moment sum rules \cite{PhysRevB.7.4300}, it is also straightforward to see that the so-called screening requirement is independent of the dimension of the space:
\begin{align}
\lim_{q\rightarrow 0 } [\epsilon_D(\qq,0)]^{-1} = 0\,.
\end{align}

\section{Conclusion and outlook} 

 In this paper we have extended the celebrated Singwi, Tosi, Land, and Sjölander (STLS) scheme for the homogeneous electron liquid (HEL) ---whose accuracy to compute the full electronic density response is comparable to Quantum Monte-Carlo--- to arbitrary integer dimensions. Our main motivation was to study the quality of the random phase approximation (RPA) results for the HEL in arbitrary integer dimensions. To that goal, we have provided new analytical formulae for the real and imaginary parts of the Lindhard polarizability and for the so-called local field correction of the STLS theory. This improves the RPA in the density-density response.  We have also discussed the un-physical predictions of the STLS method for large values of $r_s$ in arbitrary dimensions and the incapability of the scheme of fulfilling the compressibility sum rule. From our results we can conclude that the algebraic properties of the correlation energy found in Ref.~\cite{PhysRevB.102.035123} are mainly valid in the high-density limit. Furthermore, in agreement with what is known in 2 and 3 dimensions, the RPA tends to over-correlate the gas also in the high dimensions studied here. We have shown the versatility of STLS to tackle arbitrary dimensions, which can potentially shed light on the more far-reaching problems of quantum many-body systems embedded  in fractional or synthetic dimensions \cite{Kempes,PhysRevB.100.155135,PhysRevResearch.2.013044,PhysRevB.97.195101,PhysRevB.98.205116}. In fact, we have found that the compressibility sum rule behaves better when the dimension increases.   We believe the results of this paper can be a useful framework to improve our overall comprehension of Coulomb gases, and to develop a more coherent and unified dimensional approach to the correlation problem of those systems \cite{Constantin,C7CP01137G,Lewin,doi:10.1063/1.5143225}. 

\begin{acknowledgements}
We thank Robert Schlesier for helpful discussions.
We acknowledge financial support from “BiGmax”, the Max
Planck Society’s research network on big-data-driven materials science, and the European Union’s Horizon Europe Research and Innovation program  under the Marie Skłodowska-Curie grant agreement n°101065295 (C.L.B.-R.).
\end{acknowledgements}

\newpage
\onecolumngrid

\appendix

\section{Calculation of the Lindhard polarizability in $D \geq 3$} \label{app1}
To compute the Lindhard polarizability for dimensions larger than 3 we start by writing the one-particle density and the external potential as linear perturbations from equilibrium:
\begin{align}
  f(\rr,\pp, t) &= f_0(\pp) + \lambda \,\delta f (\rr,\pp, t) \,, \notag \\
  v(\rr, t) &= v_0(\rr) + \lambda \, \delta v(\rr, t) \,.
  \label{eqa1}
\end{align}
After substituting Eq.~\eqref{eqa1} in  Eq.~\ref{eq1} one obtains the following result for the induced charge density $\rho_{\textrm {ind}}(\qq,\omega)$:
\begin{align}
    \rho_{\textrm {ind}}(\qq,\omega) = \frac{\chi^0_D(\qq, \omega)}{1 - \Phi(\qq) [1 - G_D(\qq)] \chi^0(\qq, \omega)} \delta v(\qq,\omega) \,,
\end{align}
where $\chi^0_D(\qq, \omega)$ is given by:
\begin{align}
    \chi^0_D(\qq,\omega) &= -\lim_{\eta \rightarrow 0^+} \int \frac{\qq\cdot\nabla_\pp f_0(\pp)}{\omega + i \eta - \pp \cdot \qq} d^D\pp
\end{align}
By using the Taylor expansion:
$f_0(\pp\pm\tfrac{1}{2}\qq) = f_0(\pp) \pm \tfrac{1}{2} \qq \cdot \nabla_\pp f_0(\pp) + \cdots$,
we can then rewrite $\chi^0_D(\qq, \omega)$ as:
\begin{align}
    \chi^0_D(\qq,\omega) \approx \lim_{\eta \rightarrow 0^+} \int \frac{f_0(\pp-\frac{1}{2}\qq) -f_0(\pp +\frac{1}{2}\qq)}{\omega + i\eta - \pp \cdot \qq} d^D\pp \,.
    \label{eqa2}
\end{align}
This is the  Lindhard polarizability of a $D$-dimensional Fermi gas \cite{SINGWI1982177}. 

At $T = 0$ the equilibrium density reads
\begin{gather}
    f_0(\pp \pm \tfrac{1}{2}\qq) = \frac{2}{(2 \pi)^D} \Theta\left[ \tfrac{1}{2}\left( q_F^2 - \left|\pp \pm \tfrac{1}{2}\qq \right|^2 \right) \right] = \frac{2}{(2 \pi)^D} \Theta ( q_F - |\pp \pm \tfrac{1}{2}\qq|) \,.
\end{gather}
Plugging these equations and the identity
$\lim_{\eta \rightarrow 0^+} \frac{1}{x-x_0 \pm i \eta} = \mathcal{P} \frac{1}{x-x_0} \mp i \pi \delta(x-x_0)$
into Eq.~\eqref{eqa2} yields:
\begin{align}
    \chi^0_D(\qq, \omega) &= \frac{2}{(2 \pi)^D} \mathcal{P} \int \frac{\Theta (q_F - |\pp - \frac{1}{2}\qq|) - \Theta (q_F - |\pp + \frac{1}{2}\qq|)}{ \omega - \pp \cdot \qq} d^D \pp \notag \\ &- i \frac{2 \pi}{(2 \pi)^D} \int \left[\Theta \left(q_F - \left|\pp - \frac{1}{2}\qq \right| \right) - \Theta \left( q_F - \left|\pp + \frac{1}{2}\qq \right| \right) \right] \delta \left( \omega - \pp \cdot \qq \right) d^D \pp \,.
\end{align}

The real part of $\chi^0_D(\qq, \omega)$ can then be rewritten as:
\begin{align}
    \mathrm{Re} \chi^0_D (\qq, \omega) &= \frac{2}{(2 \pi)^D} \mathcal{P} \int \frac{\Theta(q_F - |\pp - \tfrac{1}{2}\qq|) \Theta(|\pp + \tfrac{1}{2}\qq|- q_F)}{ \omega - \pp \cdot \qq} d^D \pp \notag \\ & \quad - \frac{2}{(2 \pi)^D} \mathcal{P} \int \frac{\Theta(|\pp - \tfrac{1}{2}\qq| - q_F) \Theta(q_F - |\pp + \tfrac{1}{2} \qq|) }{ \omega - \pp \cdot \qq} d^D\pp \,.
\end{align}
The second integral can be rewritten with the substitution $\pp \rightarrow - \pp$ so that the expression of $\mathrm{Re} \chi^0_D (\qq, \omega)$ reduces to just one single integral:
\begin{align}
    \mathrm{Re} \chi^0_D (\qq, \omega) = \frac{2}{(2 \pi)^D} \mathcal{P} \int \Theta( q_F - |\pp - \tfrac{1}{2}\qq| ) \Theta(|\pp + \tfrac{1}{2} \qq| - q_F) \left( \frac{1}{ \omega - \pp \cdot \qq} - \frac{1}{{ \omega + \pp \cdot \qq}} \right) d^D \pp\,.
\end{align}
Utilizing the following property of the Heaviside step function: $\Theta(x) = 1 - \Theta(-x)$ and the symmetry of the integrand under the transformation $\pp \rightarrow - \pp$, $\mathrm{Re} \chi^0 (\qq, \omega)$ can be further simplified to:
\begin{align}
    \mathrm{Re} \chi^0_D (\qq, \omega) = \frac{2}{(2 \pi)^D } \mathcal{P} \int \Theta(q_F - |\pp - \tfrac{1}{2}\qq|) \left( \frac{1}{ \omega - \pp \cdot \qq} - \frac{1}{{ \omega + \pp \cdot \qq}} \right) d^D\pp \,.
\end{align}
This integral becomes dimensionless by  putting $\pp - \frac{ \qq}{2} = \kk$ and introducing the following substitutions $\qqn = \qq/q_F$, $\kkn = \kk/q_F$ and $\wn = \omega/q_F^2$:
\begin{align}
    \mathrm{Re} \chi^0_D (\qq, \omega) = \frac{2 q_F^{D-2}}{(2 \pi)^D} \mathcal{P} \int \Theta(1 - \kn) \left( \frac{1}{\wn - \qqn \cdot \kkn - \frac{\qn^2}{2}} - \frac{1}{\wn + \qqn \cdot \kkn + \frac{\qn^2}{2}} \right) d^D\kkn\,.
\end{align}
To solve this integral we choose $D$-dimensional spherical coordinates such that $\qqn$ is parallel to the last axis:
\begin{align}
    \mathrm{Re} \chi^0_D (\qq, \omega) = \frac{2 q_F^{D-2}}{(2 \pi)^D} \frac{2 \pi^{\frac{D-1}{2}}}{\Gamma \left( \frac{D-1}{2} \right)} \mathcal{P} \int_0^1 \int_0^{\pi} \left( \frac{\kn^{D-1} \sin^{D-2} \theta}{\wn - \qn \kn\cos \theta - \frac{\qn^2}{2}} - \frac{\kn^{D-1} \sin^{D-2}\theta}{\wn + \qn \kn\cos \theta + \frac{\qn^2}{2}} \right) d \theta d\kn \,,
\end{align}
where $\theta$ is the angle between $\qqn \text{ and } \kkn$. The full evaluation of this integral is quite involved. We just mention that this gives the known result for $D = 3$:
\begin{align}
    \mathrm{Re} \chi^0_{3}(\qq, \omega)
    &= \frac{ q_F}{2 \pi^2} \Bigg\{ -1 + \frac{1}{2\qn} \left[1- \left( \frac{2\wn - \qn^2}{2\qn} \right)^2 \right] \ln \left| \frac{2\qn  -\qn^2 +2\wn}{2\qn +\qn^2 - 2\wn} \right| \notag \\ & \quad- \frac{1}{2\qn} \left[1- \left( \frac{2\wn + \qn^2}{2\qn} \right)^2 \right] \ln \left| \frac{2\qn + \qn^2 +2\wn}{2\qn - \qn^2 - 2\wn} \right|  \Bigg\} \,. \label{eq26}
\end{align}
We provide the explicit results for $D = 5$ in \eqref{LPD5} and $D = 7$ in \eqref{LPD7}.

The imaginary part of the Lindhard polarizability can be computed by using the same substitutions as in the calculation of $\mathrm{Re} \chi^0_D (\qq, \omega)$:
\begin{align}
    \mathrm{Im} \chi^0_D(\qq, \omega) = - \frac{q_F^{D-2}}{(2 \pi)^{D-1}} \int [\Theta(1 - \kn) - \Theta(1-|\qqn + \kkn|)] \delta \left( \wn - \qqn \cdot \kkn - \frac{\qn^2}{2} \right) d^D\kkn \,.
\end{align}
Next we want to investigate the symmetry of $\mathrm{Im} \chi^0_D(\qq, \omega)$. By considering $\mathrm{Im} \chi^0_D(-\qq, -\omega)$ and substituting $ \kkn' = \kkn - \qqn$ we realize that $\mathrm{Im} \chi^0_D(\qq, \omega) = - \mathrm{Im} \chi^0_D(-\qq, -\omega)$. This symmetry allows us to restrict ourselves to case where $\omega > 0$. $\mathrm{Im} \chi^0_D(\qq, \omega)$ can then be rewritten for positive $\omega$ as follows:
\begin{align}
    \mathrm{Im} \chi^0_D(\qq, \omega) = - \frac{q_F^{D-2}}{(2 \pi)^{D-1}} \int \Theta(1 - \kn)\Theta(|\qqn + \kkn|-1) \delta \left( \wn - \qqn \cdot \kkn - \frac{\qn^2}{2} \right) d^D\kkn \,.
\end{align}
There are 2 cases where the integrand is different from 0:
\begin{itemize}
    \item $\left|\qn - \frac{\qn^2}{2}\right| \leq \wn \leq \qn + \frac{\qn^2}{2}$
    \item $\wn < \qn - \frac{\qn^2}{2}\,, \ \qn < 2$
\end{itemize}
The evaluation of both integrals is technically similar. Thus we just want to discuss the evaluation in the first case. The minimum value of $\kn$ in the first case is $\kn_{\textrm{min}} = |\wn/\qn - \qn/2|$. Performing the integration over all angles except the angle $\theta$ between $\qqn$ and $\kkn$ and substituting $\cos \theta = t$ leads to:
\begin{align}
    \mathrm{Im} \chi^0_D(\qq, \omega) =  - \frac{q_F^{D-2}}{(2 \pi)^{D-1}} \frac{2 \pi^{\frac{D-1}{2}}}{\Gamma \left( \tfrac{D-1}{2} \right)}\frac{1}{\qn} \int^1_{\left| \tfrac{\wn}{\qn} - \tfrac{\qn}{2} \right|} \int^1_{-1} \delta \left( \tfrac{\wn}{\qn \kn} - \tfrac{\qn}{2\kn} - t \right) \kn^{D-2} (1-t^2)^{\tfrac{D-3}{2}} dt d\kn\,.
\end{align}
To evaluate it we should put $\wn = \qn^2/2 + \qn \kn \cos\theta \Rightarrow -1 \leq \wn/\qn \kn - \qn/2\kn \leq 1$, so that the integral over $t$ can be carried out easily. We obtain in this case the following result:
\begin{align}
     \mathrm{Im} \chi^0_D(\qq, \omega) = h(D) \dfrac{1}{\qn} \left[1- \nu_{-}^2 \right]^{\tfrac{D-1}{2}} \,,
\end{align}
where we define $h(D) = q_F^{D-2}\left[2^{D-2} (D-1) \pi^{\tfrac{D-1}{2}} \Gamma \left( \tfrac{D-1}{2} \right)\right]^{-1}$ and $\nu_{\pm} = \wn/\qn \pm \qn/2$. 

In the second case, $k_{\textrm{min}} = \sqrt{1 - 2\omega}$ and we obtain the following result:
\begin{align}
    \mathrm{Im} \chi^0_D(\qq, \omega) = h(D) \dfrac{1}{\qn} \left[ \left(1- \nu_{-}^2 \right)^{\frac{D-1}{2}} - \left( 1-  \nu_{+}^2\right)^{\frac{D-1}{2}}\right]\,.
\end{align}
Eventually $\mathrm{Im} \chi^0_D(\qq, \omega)$ reads as in Eq.~\eqref{impart}.

\section{The local field correction in the Hartree-Fock approximation} \label{app2}

\begin{figure}[!b]
    \centering
    \includegraphics[width = 0.43\linewidth]{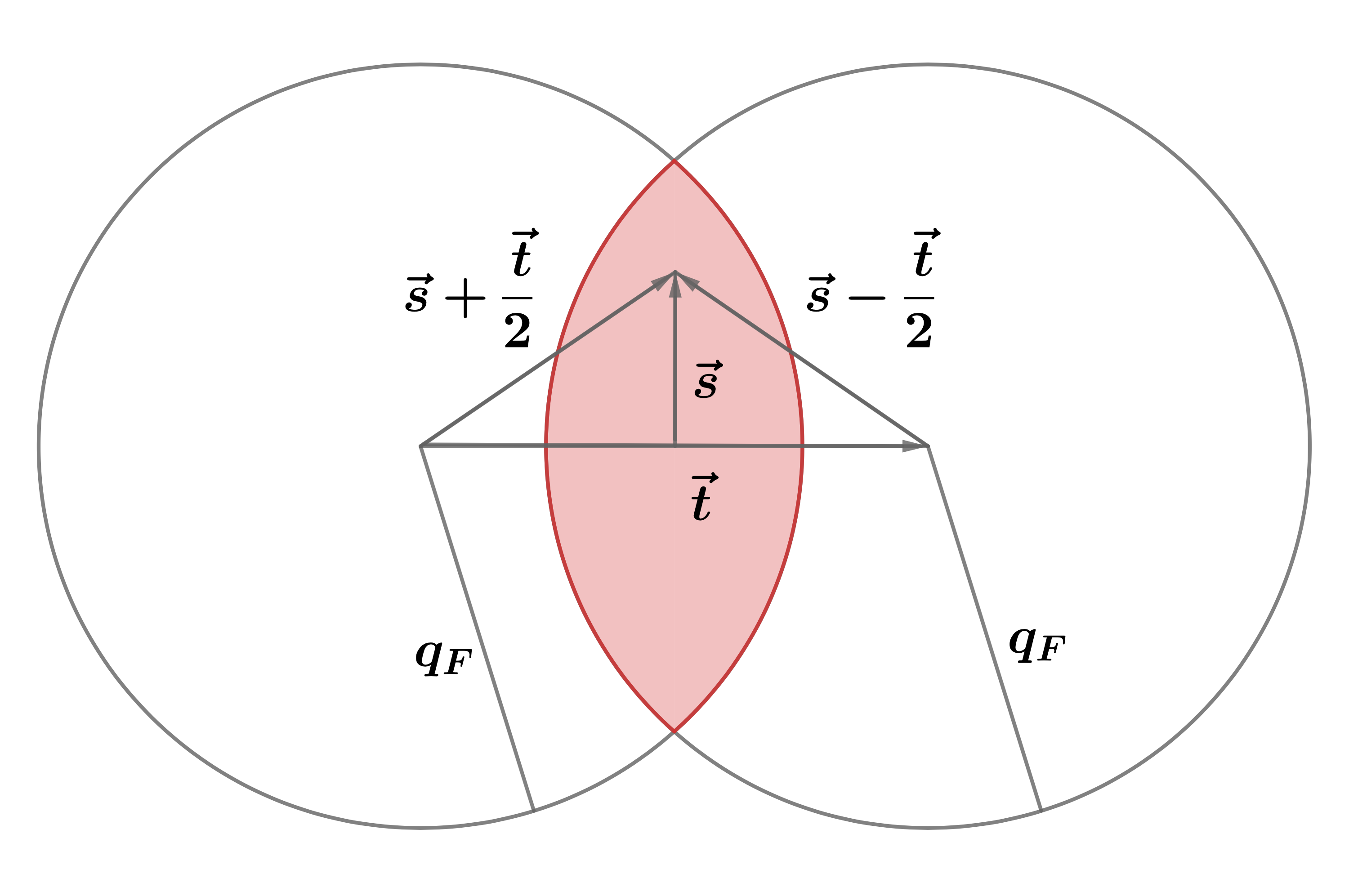}
    \caption{Integration region of $G_D^{\textrm{HF}} (\qq)$.}
    \label{overlap}
\end{figure}

After the substitution $\vs = (\kk + \kk')/2 \text{ and } \vt = \kk - \kk'$ Eq.~\eqref{eqGHF} becomes:
\begin{align}
    G_D^{\textrm{HF}}(\qq) = \frac{2 q^{D-3}}{(2 \pi)^{2D} n^2} \int_{|\vs + \frac{\vt}{2}| \leq q_F} \int_{|\vs - \frac{\vt}{2}| \leq q_F} \frac{\qq \cdot ( \qq + \vt )}{| \qq + \vt |^{D-1}} d\vt d\vs \,.
\end{align}
Because the integrand only depends on $\qq$ and $\vt$, we can perform the integration over $\vs$. This gives us the volume of the integration region and the integral becomes:
\begin{align}
    G_D^{\textrm{HF}}(\qq) = \frac{2 q^{D-3}}{(2 \pi)^{2D} n^2} \int d\vt \frac{\qq \cdot ( \qq + \vt )}{| \qq + \vt |^{D-1}} \int d\vs \Theta(q_F - |\vs + \tfrac{\vt}{2}|) \Theta(q_F - |\vs - \tfrac{\vt}{2}|) \,.
\end{align}
Thus we need to identify the region in hyperspace in which we are integrating. The region after the substitution is mathematically defined by $|\vs + \vt/2| \leq q_F \text{ and } |\vs - \vt/2| \leq q_F$ and therefore can be seen as the overlap region of two Fermi spheres in hyperspace (as shown by the colored region in Fig.~\ref{overlap}). The centers of the spheres sit at a distance $|\vt|$ from each other, the origin of $\vs$ is the midpoint of the connecting line of two centers. 
The integration over $\vs$ is the volume of this overlap region, which is the combined volumes of two identical hyperspherical caps. The volume of a $D$-dimensional hyperspherical cap was already calculated by Li \cite{Li2011ConciseFF}:
\begin{align}
    \mathcal{V}_D = \frac{1}{2} V_D(r) \textrm{I}_{\sin^2\phi} \left(\frac{D+1}{2}, \frac{1}{2} \right) \notag \,,
\end{align}
where $V_D(r)$ is the volume of a hypersphere with radius $r$ in $D$-dimensional space, $\phi$ is the angle between a vector and the positive
$D^{\mathrm{th}}$-axis of the sphere and $\textrm{I}(x,y)$ is the regularized incomplete beta function. In our case it can easily be verified that $\textrm{sin}^2 \phi = 1- \frac{t^2}{4q_F^2}$ and $V_D(r) = \pi^{D/2} q_F^D/\Gamma \left(\tfrac{D}{2} +1 \right)$. As a result, $G^D_{\textrm{HF}}(\qq)$ becomes:
\begin{align}
    G^{\textrm{HF}}_D(\qq) = \frac{2 q^{D-3}}{(2 \pi)^{2D} n^2} \frac{\pi^{D/2}}{\Gamma \left(\frac{D}{2} +1 \right)} q_F^D \int \textrm{I}_{1-\frac{t^2}{4q_F^2}} \left(\frac{D+1}{2}, \frac{1}{2} \right) \frac{\qq \cdot ( \qq + \vt )}{| \qq + \vt |^{D-1}} d\vt \,.
\end{align}
By changing to $D$-dimensional spherical coordinates such that $\qq$ is parallel to the $D^{\mathrm{th}}$-axis and using the relation between the density and the Fermi wavelength $q_F$ we obtain Eq.~\eqref{eqGHF2}. For some cases this calculation can be done analytically. Here we provide the result for the $5D$ case:
\begin{align}
    G^{\textrm{HF}}_{5}(q) &= \frac{-5}{4928} \left(\frac{q}{q_F}\right)^8 + \frac{25}{1232} \left(\frac{q}{q_F}\right)^6 + \frac{1775}{7392} \left(\frac{q}{q_F}\right)^4 - \frac{5}{66} \left(\frac{q}{q_F}\right)^2 + \frac{25}{154} \notag \\
    & + \left[ \frac{-15}{128}\left(\frac{q}{q_F}\right)^5 + \frac{15}{224}\left(\frac{q}{q_F}\right)^3 + \frac{5}{56}\left(\frac{q}{q_F}\right) - \frac{25}{154} \left(\frac{q_F}{q}\right) \right] \ln \left| \frac{q + 2q_F}{q - 2q_F} \right| \notag \\
     & + \left[ \frac{-5}{19712}\left(\frac{q}{q_F}\right)^{10} + \frac{5}{896}\left(\frac{q}{q_F}\right)^8 - \frac{15}{224}\left(\frac{q}{q_F}\right)^6 \right]\ln \left| \frac{q^2 - 4q_F^2}{q^2} \right| \,.
\end{align}


\section{Compressibility sum rule}
\label{appCSR}

By applying the compressibility sum rule (see, for instance, \cite{giuliani_vignale_2005}) one can rewrite the dielectric function in the limit $q \rightarrow 0$ at $\omega = 0$ as follows:
\begin{align}
    \lim_{q \rightarrow 0} \epsilon_D(q, 0) = 1 + \Phi(q) n^2 \kappa = 1+ \frac{(4 \pi)^{\frac{D-1}{2}} \Gamma\left( \frac{D-1}{2} \right)}{q^{D-1}} \frac{D r_s^2 n}{\alpha_D^2} \frac{\kappa}{\kappa_f} \,,
\end{align}
where $\kappa_f = \frac{D r_s^2}{n\alpha_D^2}$ is the free compressibility. By defining the $D$-dimensional Thomas-Fermi wavelength as:
\begin{align*}
    q_{\mathrm{TF}} = \left(\frac{(4 \pi)^{\frac{D-1}{2}} \Gamma\left( \frac{D-1}{2} \right) D r_s^2 n}{\alpha_D^2} \right)^{\tfrac{1}{D-1}}
\end{align*}
we obtain Eq.~\eqref{compren}. In 3D this amounts to the well-known result
$$   
\lim_{q\rightarrow 0 } \epsilon_3(\qq,\omega) = 1+ \frac{q_{\rm TF}^2}{q^2} \frac{\kappa}{\kappa_f}\,.
$$
Finally, by applying the sum rules to the Lindhard polarizability (i.e.,
$
    \lim_{q \rightarrow 0} \chi_0(q, 0) = -n^2 \kappa_f  \,,
$)
we obtain the dielectric function in the STLS theory (Eq.~\eqref{eq4}) in the limit $q \rightarrow 0$:
\begin{align}
   \lim_{q\rightarrow 0 }  \epsilon_D(\qq,0) = 1 + \frac{\left( \tfrac{q_{\mathrm{TF}}}{q} \right)^{D-1}}{1 -  \gamma_D \left(\tfrac{q_{\mathrm{TF}}}{q_F} \right)^{D-1}} \,,
    \label{eq4b}
\end{align}
where $\gamma_D = -\frac{1}{2q_{\mathrm{F}}} \int^\infty_0 [S_D(q)-1]dq$. From Eq.~\ref{compren} one eventually finds:
\begin{align}
    \frac{\kappa_f}{\kappa} = 1 - \gamma_D \left(\frac{q_{\mathrm{TF}}}{q_F} \right)^{D-1} \,.
\end{align}

\twocolumngrid

\bibliography{Refs2}

\end{document}